\documentclass[
 reprint,
amsmath,amssymb,aps,pre
]{revtex4-1}
\newcommand\numberthis{\addtocounter{equation}{1}\tag{\theequation}}
\usepackage{graphicx}
\usepackage{dcolumn}
\usepackage{bm}
\usepackage[normalem]{ulem}

\begin{document}
\title{Quantum Ising model on the frustrated square lattice} 
 \author{N. Kellermann}
 \author{M. Schmidt}
 \email{mschmidt@mail.ufsm.br}
  \affiliation{Departamento de F\'isica, Universidade Federal de Santa Maria, 97105-900 Santa Maria, RS, Brazil}
\author{F. M. Zimmer}\email{fabio.zimmer@ufms.br}%
\affiliation{%
 Instituto de F\'ísica, Universidade Federal de Mato Grosso do Sul, 79070-900 Campo Grande, MS, Brazil
}%


\begin{abstract}
We investigate the role of a transverse field on the Ising square antiferromagnet with first-($J_1$) and second-($J_2$) neighbor interactions. Using a cluster mean-field approach, we provide a telltale characterization of the frustration effects on the phase boundaries and entropy accumulation process emerging from the interplay between quantum and thermal fluctuations. We found that the paramagnetic (PM) and antiferromagnetic phases are separated by continuous phase transitions. On the other hand, continuous and discontinuous phase transitions, as well as tricriticality, are observed in the phase boundaries between PM and superantiferromagnetic phases. A rich scenario arises when a discontinuous phase transition occurs in the classical limit while quantum fluctuations recover criticality.
We also find that the entropy accumulation process predicted to occur at temperatures close to the quantum critical point can be enhanced by frustration. Our results provide a description for the phase boundaries and entropy behavior that can help to identify the ratio $J_2/J_1$ in possible experimental realizations of the quantum $J_1$-$J_2$ Ising antiferromagnet.

\end{abstract}

\pacs{Valid PACS appear here}
\maketitle

\section{Introduction}

Quantum phase transitions arise when tuning a non-thermal parameter introduces a competition between ground state phases. Despite these phenomena occur only at zero temperature, when quantum fluctuations drive a continuous phase transition, underlying signatures of  quantum criticality can be observed even at finite temperatures \cite{Vojta_QPT_Review}. 
This scenario is even richer in the presence of  frustration - the inability to simultaneously satisfy all the interactions - that can play a significant role in the phase transitions as well as the system entropy \cite{Lacroix_book}.
In addition, entropy accumulation is expected at finite temperatures in the proximity of a quantum critical point \cite{Nature_entropy_landscape}. Therefore, a worthwhile subject concerns the subtleties of systems hosting frustration and both thermal and quantum fluctuations.  

From the theoretical point of view, the transverse Ising model on the square lattice is the simplest model to 
exhibit both classical and quantum phase transitions. 
For instance, when first-neighbor ($J_1$) antiferromagnetic (AF) interactions are considered, increasing a transverse magnetic field can change the ground-state of this model from a N\'eel  AF long-range order to a polarized paramagnetic (PM) state at a quantum critical point \cite{PhysRevB.92.174419}. 
By considering also second-neighbor antiferromagnetic interactions ($J_2$), the model becomes the so called $J_1$-$J_2$ Ising model, in which frustration  can be introduced by tuning $g \equiv J_2/J_1$. 
In this case,  the AF ground-state persists for $g<0.5$, while a superantiferromagnetic (SAF) ground-state (characterized by alternated ferromagnetic rows or columns) occurs for $g>0.5$ \cite{PhysRevB.87.144406}. 
Moreover, in the absence of transverse fields, the nature of the phase transitions has been clarified only recently, indicating that the model shows continuous and discontinuous phase transitions as well as tricriticality  \cite{PhysRevE.98.022123, PhysRevB.87.144406, PhysRevE.97.022124, PhysRevB.84.174407, cvm_Moran, pd_mc_epjb, GUERRERO2017596, PhysRevLett.108.045702,  PhysRevB.86.134410}.   
In particular, the thermally driven phase transitions between SAF and PM states are discontinuous for $0.5<g<g^*$ and continuous for $g>g^*$, where $g^*$ locates the tricriticality.

The phase diagram in the classical limit indicates that the quantum $J_1$-$J_2$ Ising model can show a variety of interesting phenomena. A natural question that arises is whether the quantum fluctuations can change the nature of the phase transitions. 
For instance, several systems are known for exhibiting phase transitions that are continuous when driven by thermal fluctuations and become discontinuous when the quantum fluctuations are increased \cite{RPP_Vojta_2018}.
On the other hand, it has been pointed that some ferroelectrics systems show quantum criticality, despite the thermally driven transitions are discontinuous \cite{RPP_Chandra_2017}. This has motivated recent theoretical efforts, suggesting the possibility of a quantum annealed criticality in this class of systems \cite{2018arXiv180511771C}. 
Therefore, the study of transverse field effects in the $J_1$-$J_2$ Ising model can also contribute to the understanding of the role of quantum fluctuations and competing interactions on the phase transitions.

Despite several studies have addressed the magnetic behavior of the classical $J_1$-$J_2$ Ising model, only few efforts have been made to understand the effect of transverse fields in this model \cite{PhysRevB.92.174419,PhysRevB.94.214419,PhysRevE.97.022124}.
Recently, a cluster operator approach was used to describe the highly frustrated limit ($g=0.5$) of the quantum $J_1$-$J_2$ Ising model at zero temperature \cite{PhysRevB.94.214419}. It was found 
that a string valence-bond-solid state occurs at weak transverse fields. Very recently, a single-site effective field theory was proposed to analyze the quantum $J_1$-$J_2$ Ising model \cite{PhysRevE.97.022124}. The authors reported  discontinuous phase transitions between the AF and PM states.
However, several approaches indicated that only continuous phase transitions are expected between the AF and PM phases in the classical regime \cite{pd_mc_epjb, cvm_Moran, PhysRevE.98.022123}. 
Moreover, the analysis in Ref. \cite{PhysRevE.97.022124} is constrained to $g \leq 0.5$ due to the inability of the technique in reproducing the expected ground-state ordering for $g > 0.5$.

Therefore, the effect of transverse fields for a larger range of $g$ in this model still lacks a proper description. In the present work we address this issue, by analyzing the interplay of thermal and quantum fluctuations in the $J_1$-$J_2$ Ising model within the cluster mean-field (CMF) theory. In this technique, the intracluster interactions are incorporated exactly and the couplings between clusters are approximated by mean fields. The CMF framework has been considered in many recent investigations of spin models \cite{PhysRevLett.114.027201, PhysRevB.96.014431, PhysRevB.87.144406, PhysRevB.91.174424, PhysRevB.98.104429, PhysRevE.98.022123, JPCM_j1_j2_Heisenberg} in which frustration and quantum fluctuations are often present. In particular, this method was applied to the Ising \cite{PhysRevB.87.144406} and Heisenberg \cite{JPCM_j1_j2_Heisenberg} versions of the $J_1$-$J_2$ model, providing a description for the nature of the magnetic phase boundaries in agreement with state-of-art numerical and analytical calculations. In addition, this approach goes beyond single-site approximations, allowing to incorporate short-range correlations and improving the description of thermodynamic quantities \cite{PhysRevB.98.104429}. It means that the technique can provide insights into the 
frustration outcome on the entropy accumulation in the verge of quantum phase transitions \cite{Nature_entropy_landscape}, a topic that is also addressed in the present work.

The paper is organized as follows. The model and the CMF theory are described in Sec. \ref{met}. In Sec. \ref{res}, we present the results obtained, discussing the role of frustration and transverse field in the phase diagrams  of the model. In addition, our findings indicate that the entropy accumulation phenomenon can be enhanced by frustration. Finally, in Sec. \ref{conc}, we summarize the paper and present the conclusion.

\section{Model and Method}\label{met}
The transverse Ising model is given by the Hamiltonian
\begin{equation}
    H=-\sum_{i,j}J_{ij}\sigma^z_i \sigma^z_j - \Gamma \sum_{i} \sigma^x_i \label{hamiltonian}
\end{equation}
where $\sigma^l_i$ is the $l$ component of the Pauli matrices at site $i$. $J_{ij}$ and $\Gamma$ are the exchange coupling between pairs of spins and the transverse magnetic field, respectively.
We assume AF first-neighbor ($J_1<0$) and  second-neighbor ($J_2<0$) exchange interactions. 

An exact solution for the model is still unavailable at a nonzero $\Gamma$ and/or $J_2$. In this case, the CMF theory can provide a reputable framework. In fact, CMF approaches have been adopted in several recent studies in condensed matter physics and statistical mechanics \cite{PhysRevLett.114.027201, PhysRevB.96.014431, PhysRevB.87.144406, PhysRevB.91.174424, PhysRevB.98.104429, PhysRevE.98.022123, JPCM_j1_j2_Heisenberg}. 
The technique consists in dividing the lattice in $N_{cl}$ equivalent clusters with $n_s$ sites each. The couplings inside the clusters are evaluated by exact diagonalization and the intercluster interactions are replaced by the mean-field approximation: $\sigma^l_i \sigma^l_j \approx \sigma^l_i m^l_j+  m^l_i \sigma^l_j - m^l_i m^l_j$. 
Minimization of free-energy leads to the standard mean-field self-consistent equations $m^l_i = \langle \sigma^l_i  \rangle$, where $\langle \cdots \rangle$ accounts for the thermodynamic average. 

One of the main motivations for choosing this approach is the qualitatively correct results provided by it 
in the classical limit ($\Gamma=0$) of the $J_1$-$J_2$ Ising model. Despite it fails by showing an ordering temperature in the highly frustrated point ($g=0.5$), the technique leads to accurate estimates of $g^*$, providing a good description for the nature of the phase transitions. 
For example, for clusters with four sites, the CMF theory yields $g^*= 0.66$, which agrees very well with the most recent Monte Carlo results ($g^*_{MC}=0.67$) \cite{PhysRevLett.108.045702, PhysRevB.86.134410}. 
It is worth to mention that this CMF result is also robust under cluster size increase, as shown for a 16-site cluster approximation \cite{PhysRevE.98.022123}. 
It indicates that the nature of the phase transitions is correctly incorporated within a four-site approximation, allowing to explore transverse field effects without appealing to the exact diagonalization of larger matrices, which would become computationally expensive taking into account the self-consistent procedure. 
In addition, intermediary cluster sizes ($4<n_s<16$) can imply in clusters that are incompatible with the SAF state \cite{PhysRevE.91.032145, PhysRevE.98.022123}. Therefore, in the present work we adopt the four-site CMF approach, which is described in the following.

\begin{figure}
 \includegraphics[width=0.67\columnwidth]{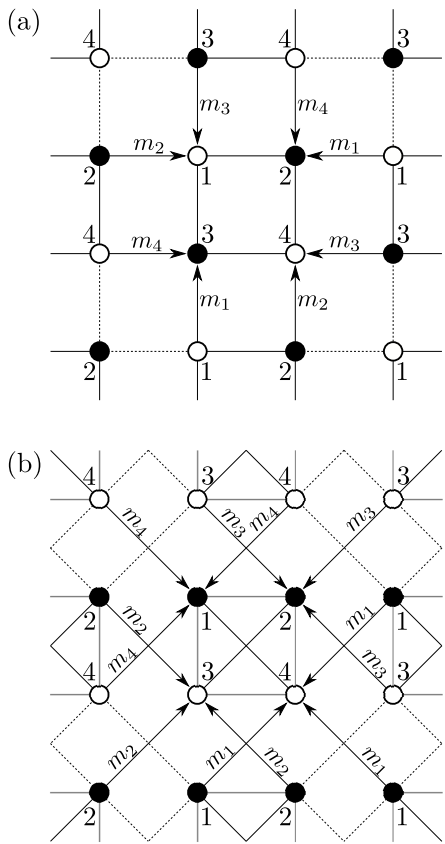} 
 \caption{Schematic view of the four-site cluster mean-field approximation. First-neighbor and second-neighbor interactions, including mean-fields, are depicted in panels (a) and (b), respectively. The mean fields acting on the central cluster are denoted by arrows and the intracluster couplings are represented by solid lines. For clarity, the first-neighbor intracluster couplings are showed in (b). Solid and open circles denote different ordering structures, in which (a) and (b) represent AF and SAF configurations, respectively. }
 \label{cmf}
\end{figure}

The fundamental advantage of the CMF procedure is to decouple the clusters in such a way that the many-body problem (see Eq. (\ref{hamiltonian})) becomes a single-cluster one. In the four-site approximation,
the CMF Hamiltonian (see Fig. \ref{cmf}) is given by
\begin{equation}
 H_{\textrm{MF}}=H_{\textrm{intra}}+H'-\Gamma \sum_{i}^{n_s=4} \sigma^{x}_i \end{equation}
where the intracluster term is given by
\begin{equation}
 H_{\textrm{intra}}=-J_{1}(\sigma^z_1+\sigma^z_4)( \sigma^z_2+ \sigma^z_3) -J_{2}(\sigma^z_1 \sigma^z_4+\sigma^z_2 \sigma^z_3),
\end{equation}
and the mean-field contribution can be written as 
\begin{align*}
H'=&-J_{1}[(\sigma^z_1+\sigma^z_4)( m^z_2+ m^z_3)+(\sigma^z_2+\sigma^z_3)( m^z_1+ m^z_4)] 
 \\&+J_1(m^z_2 + m^z_3)( m^z_1 + m^z_4)-3J_{2}[\sigma^z_1 m^z_4+\sigma^z_2 m^z_3 
 \\ &+\sigma^z_3 m^z_2+\sigma^z_4 m^z_1]+ 3 J_2(m^z_1 m^z_4 + m^z_2m^z_3). \numberthis \label{mf_ham}
\end{align*}
After evaluating the on site magnetizations, $m^z_i$, one can compute the thermodynamics from the free-energy per spin,
\begin{equation}
f=-\frac{k_B T}{n_s}\ln{Z} 
\label{free_energy}    
\end{equation}
where  $T$ is the temperature, $k_B$ is the Boltzmann constant - assumed to be one - and the partition function is given by $Z=\textrm{Tr} \, \textrm{e}^{-\beta H_{\textrm{MF}}}$, with $\beta=1/T$. The entropy per site is then given by
\begin{equation}
\frac{S}{n_s}=\frac{u-f}{T},    
\end{equation}
where 
\begin{equation}
 u=\langle  H_{\textrm{MF}} \rangle/n_s = \frac{\textrm{Tr} ~ H_{\textrm{MF}}\, \textrm{e}^{-\beta H_{\textrm{MF}}}}{n_s Z}  
\end{equation}
is the internal energy per spin. 

The AF and SAF long-range orders (see Fig. \ref{cmf}) are  described by the order parameters 
\begin{equation}
m_{\textrm{AF}}=m^z_1-m^z_2-m^z_3+m^z_4 \neq 0
\end{equation}
and
\begin{equation}
m_{\textrm{SAF}}=|m^z_1-m^z_4|+|m^z_2-m^z_3| \neq 0,
\end{equation}
respectively. The PM state occurs when $m_{AF}=m_{SAF}=0$ and the location of discontinuous phase transitions are done by comparing the free energies (Eq. \ref{free_energy}) of the different solutions.

\section{Results}\label{res}

\begin{figure}
 \includegraphics[width=1.\columnwidth]{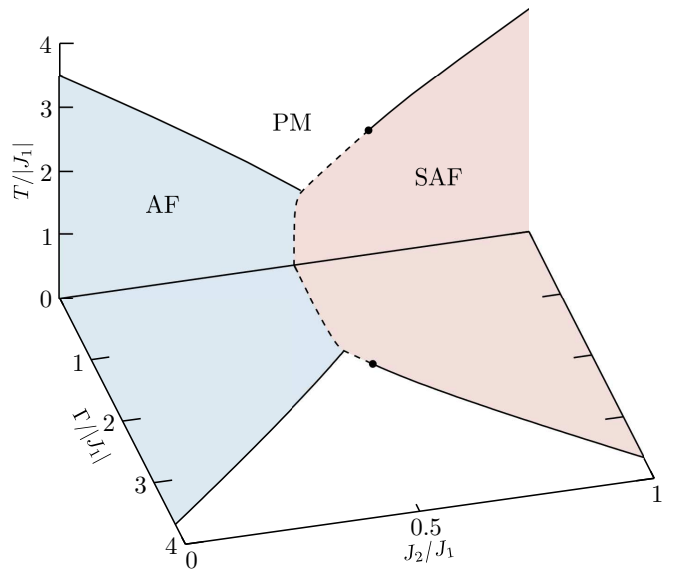} 
\caption{ Global phase diagram for the frustrated square lattice in a transverse field. Solid and dashed lines indicate continuous and discontinuous 
phase transitions, respectively. The black circles denote the tricritical points. }
 \label{global_pd}
\end{figure}

In Fig. \ref{global_pd}, we present the global phase diagram of the frustrated square lattice in a transverse field. In the absence of transverse fields, when $J_2/J_1 \to 0.5$, a reduction of the phase transition temperature is observed. While the PM/AF phase boundary is only continuous, the transitions between AF and SAF phases are discontinuous. Moreover, lines of continuous and discontinuous phase transitions between the PM and SAF phases are separated by a tricritical point. These results recover those of Refs. \cite{PhysRevB.87.144406, PhysRevE.98.022123}. 
In particular, clusters with $n_s=4$ and $n_s=16$ were considered in Ref. \cite{PhysRevE.98.022123}, but the nature of the phase transitions remained unchanged with the increase of cluster size.

It is worth to stress that the nature of the phase transitions in this model is a topic of current debate. For instance, an effective field-theory study using clusters of up to 9 sites indicated the presence of a discontinuous  phase transition between PM and AF states \cite{PhysRevE.91.032145}. However,  our results  indicating only continuous phase transitions for the PM/AF phase boundary are in agreement with Monte Carlo simulations \cite{pd_mc_epjb} and cluster variation method calculations \cite{cvm_Moran}. 
The nature of the transitions between the AF and SAF states is also in good agreement with very recent numerical results for this model.
In particular, our approach indicates that tricriticality occurs at $g*=0.66$ \cite{PhysRevB.87.144406, PhysRevE.98.022123}, which is much closer to the Monte Carlo simulations ($g^*_{\textrm{MC}} \approx 0.67$) \cite{PhysRevB.86.134410, PhysRevLett.108.045702} than the effective field theory predictions ($g^*_{\textrm{EFT}} = 0.97$) for the largest cluster considered in Ref. \cite{PhysRevE.91.032145}. 
It reinforces that the present approach incorporates important geometrical features of the model, providing an appropriate starting point to study the role of transverse field.
\begin{figure}
 \includegraphics[width=1.\columnwidth]{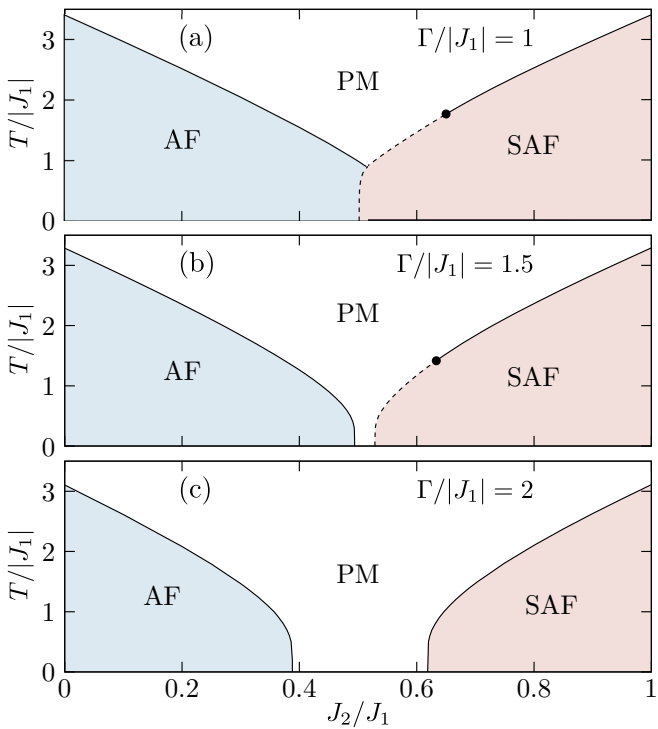} 
 \caption{ Phase diagrams for several values of $\Gamma$. It is used the same convention as in Fig. \ref{global_pd}.  }
 \label{pd_several_gamma}
\end{figure}

Now, we discuss quantum fluctuation effects.
A weak transverse field leads to qualitatively the same $T/|J_1|$ {\it versus} $J_2/J_1$ phase diagram obtained in the classical limit, only reducing the phase transition temperature, as shown in Fig. \ref{pd_several_gamma}(a). However, a strong enough $\Gamma$ leads to the onset of a PM state between AF and SAF phases even at zero temperature (see Figs. \ref{pd_several_gamma}(b)-(c)).
An interesting result is that, for a certain range of the transverse field, tuning $g$ leads to the observation of phase transitions of different natures in the ground-state. For instance, when $\Gamma/|J_1|=1.5$, PM/AF phase transitions are continuous while the PM/SAF ones are discontinuous at low temperatures.
Another scenario arises for larger transverse fields, in which only continuous phase transitions are observed, as shown for $\Gamma=2$. For sufficiently large $\Gamma$ the AF state disappears and the SAF phase is observed only for a large enough $g$. In particular, the evaluated critical transverse field for $J_2=0$ is $\Gamma_c/|J_1|=3.678$ (see the zero-temperature plane of Fig. \ref{global_pd}).

\begin{figure}
 \includegraphics[width=1.\columnwidth]{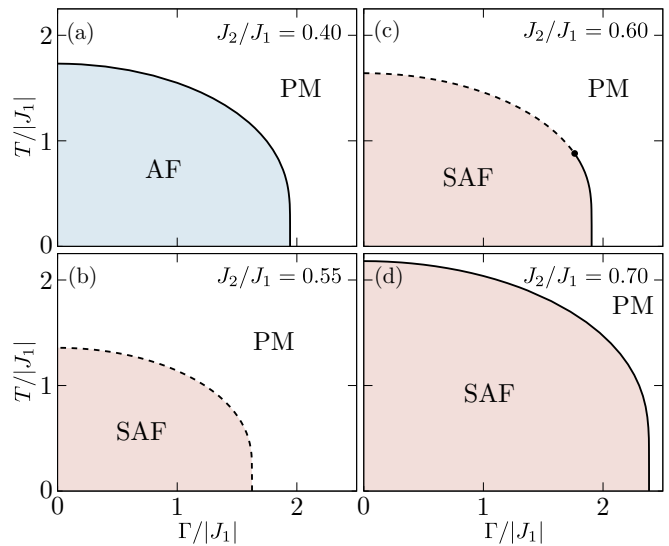} 
 \caption{Phase diagrams for different coupling ratios $J_2/J_1$. It is used the same convention as in Fig. \ref{global_pd}. }
 \label{pd_several_j2}
\end{figure}

Tuning $g$ in materials can be an intricate task. A more reasonable scenario is to realize a system with a certain $g$ and then evaluate the effect of a transverse field. Our results indicate that depending on the ratio between first-neighbor and second-neighbor interactions, four different types of $T \times \Gamma$ phase diagrams can be obtained.
We can discuss these different scenarios arising from the interplay of quantum and thermal fluctuations by analyzing the results shown in Fig. \ref{pd_several_j2}.
The phase transitions between PM and AF states remain continuous at any $\Gamma$, as shown in Fig. \ref{pd_several_j2}(a) for $J_2/J_1=0.4$. This indicates that the continuous nature of the PM/AF phase transitions is robust against quantum fluctuations.
This robustness seems to also occur for the $J_1$-$J_2$ Heisenberg model, for which recent results indicate that PM and AF states are separated by a continuous phase transition \cite{JPCM_j1_j2_Heisenberg,PhysRevB.86.024424}.

A richer scenario arises in the PM/SAF phase boundary due to the difference between the quantum tricritical point $g^*_Q=0.56$ and the thermally driven one $g^*$.
For instance, at $J_2/J_1=0.55$ the PM/SAF phase boundary is entirely discontinuous (Fig. \ref{pd_several_j2}(b)). A particularly interesting phenomena occurs for a certain range of $g$, in which the transitions become continuous when the transverse field is increased (see Fig. \ref{pd_several_j2}(c)). Finally, an entirely continuous phase boundary is observed at sufficiently larger $g$, as illustrated for $g=0.70$, in Fig. \ref{pd_several_j2}(d).

\begin{figure*}[ht]
 \includegraphics[width=0.98\textwidth]{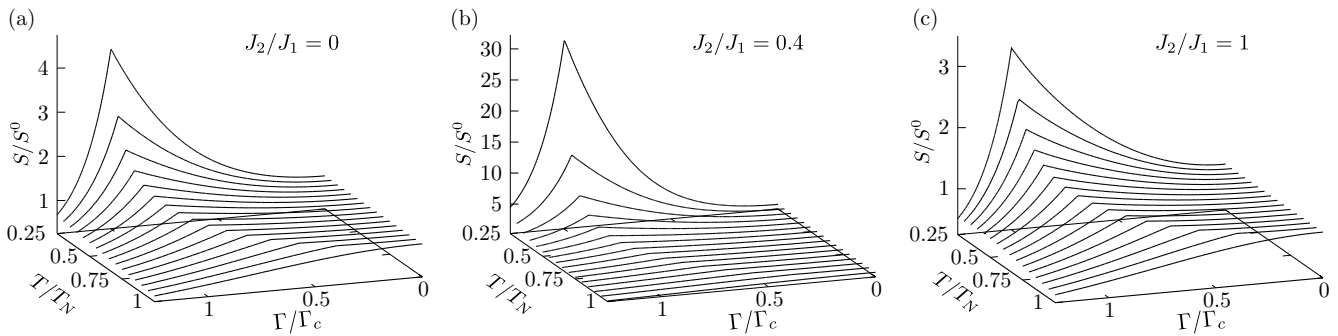} 
  \caption{Normalized entropy ($S/S^0$) as a function of transverse field ($\Gamma / \Gamma_c$) for several values of temperature ($T/T_N$) and (a) $J_2/J_1=0$, (b) $J_2/J_1=0.4$ (c) $J_2/J_1=1$, 
 where $S_0=S(T,J_2/J_1,\Gamma=0)$. 
 We analyze the temperature range from $0.25 T_N$ to $T_N$ for each $J_2/J_1$. }
 \label{fig4}
\end{figure*}

To the best of our knowledge, the only available analysis of the transverse field effects on the nature of the phase boundaries of the present model was done only recently, within a single-site effective field theory \cite{PhysRevE.97.022124}. In particular, the authors found discontinuous phase transitions even between AF and PM states, with the ground-state tricriticality occurring at smaller values of $J_2/J_1$ when compared to the thermally driven criticality at zero fields.  It suggests that transverse fields lead to discontinuous phase transitions in the PM/AF phase boundary.  Our findings indicate that the opposite occurs in the PM/SAF phase boundary, which brings an important issue on the effect of transverse fields in this model. In addition,
our results indicate that experimental realizations of the present model can host a version of the recently proposed quantum annealed criticality \cite{2018arXiv180511771C}. 
In this case, it is claimed that compressible systems can exhibit phase transitions that are discontinuous in the classical limit and continuous when these are induced by pressure at zero temperature. In our case, for a system with an intermediary level of frustration ($g^*_Q<g<g^*$), one can observe discontinuous phase transitions in the absence of quantum fluctuations and criticality driven by zero-point fluctuations. Analogous phenomena have been observed in many materials \cite{RPP_Chandra_2017}, suggesting that our results can provide an interesting additional mechanism, based on competing interactions, for quantum annealed criticality.

Underlying signatures of quantum phase transitions are often observed in the thermodynamics properties even at finite temperatures \cite{Vojta_QPT_Review,PhysRevB.92.180404,PhysRevB.92.205123}. For instance, quantum criticality is usually associated with accumulation of entropy at low but nonzero temperatures \cite{PhysRevB.97.245127,Nature_entropy_landscape}. 
This phenomenon can be observed for the present model in Fig. \ref{fig4}, where the entropy ($S$) is divided by the zero-field entropy ($S^0$) for a given $T/|J_1|$ and $J_2/J_1$.
It means that $S/S^0=1$ in the absence of the transverse field for each temperature and coupling ratio.  
As a consequence, Fig. \ref{fig4} makes clear the effects of quantum fluctuations on the entropic content.
For instance, the accumulation of entropy can be observed in the proximity of the phase boundaries, with the maximum of $S/S^0$ occurring at the phase transition. Moreover, this maximum is enhanced as temperature is lowered, suggesting a divergence at $T\to 0$.

We highlight that the effect of frustration on the process of entropy accumulation can be examined by comparing panels in Fig.  \ref{fig4}, where only continuous phase transitions are depicted.  
Despite the three panels exhibit results in qualitative agreement,  panel (b) shows higher values of $S/S^0$ in the proximity of the phase transition at low temperatures when compared with panels (a) and (c). It indicates that higher degrees of frustration ($J_2/J_1=0.4$ in panel (b)) can lead to an enhancement of the entropy accumulation phenomena. These results suggest that a more neat signature of quantum criticality can be observed in the entropy landscape of materials that exhibit a higher degree of frustration.

\section{Conclusion}\label{conc}
We study transverse field effects in the $J_1$-$J_2$ Ising antiferromagnet within a CMF approach. This technique allowed us to evaluate phase boundaries in the full range of parameters $g\equiv J_2/J_1$, $T$ and $\Gamma$, providing a global phase diagram for the model (see Fig. \ref{global_pd}).
By tuning $g$ at intermediary transverse fields, we found a low temperature PM state separating the AF and SAF long-range orders for $g\approx0.5$.
While the phase transitions between PM and AF states are always continuous, the nature of the PM/SAF phase boundaries show a strongly dependence with $\Gamma$ and $T$. 
In particular, our findings indicate that the model is a candidate to exhibit a version of the recently proposed quantum annealed criticality \cite{2018arXiv180511771C}. 
We notice that for certain values of $g$, the PM/SAF phase transition is discontinuous at $\Gamma=0$, but a continuous quantum phase transition is found by tuning $\Gamma$ at zero temperature.
This brings a possible new mechanism for the quantum annealed criticality:  the presence of competing interactions. 
Moreover, our analysis of the entropy behavior indicate that the competitive scenario related to frustration can lead to an enhanced entropy accumulation process in the verge of the critical point.

We hope that the present work motivates further studies of the $J_1$-$J_2$ quantum Ising model. In particular, the criticality induced by quantum fluctuations deserves further attention. 
A starting point would be to verify the existence and location of $g^*_Q$.  Other interesting question concerns the presence of  disorder, which cannot be completely avoided in physical systems. Since disorder can modify the nature of phase transitions \cite{RevModPhys.88.025006}, its role in the phase boundaries can be relevant for the quantum annealed criticality. 
In addition, the combination of frustration and bond disorder can lead this model to exhibit spin-glass freezing \cite{PhysRevB.97.224419,1402-4896-90-2-025809,PhysRevE.89.022120}. Therefore, the presence of disorder in this quantum model with competing interactions can provide a rich phenomenology. Finally, further  thermodynamic analysis can also confirm the entropy accumulation enhancement driven by frustration in this model.

\section*{Acknowledgments}
This work was supported by the Brazilian agencies Conselho Nacional de Desenvolvimento Cient\'ifico e Tecnol\'ogico (CNPq) and Coordena\c{c}\~ao de Aperfei\c{c}oamento de Pessoal de N\'ivel Superior (Capes).

\bibliography{references}

\end{document}